\documentclass[preprint,12pt]{elsarticle}

\usepackage{graphicx}
\usepackage{amssymb}
\usepackage{lineno}

\journal{Sensors and Actuators A: Physical}

\begin{document}

\begin{frontmatter}


% \title{Low current Hall Effect Sensor}

\title{Low current Hall Effect Sensor}
% \author{Dima Cheskis\corref{cor1}\fnref{label1}}
 \author{Yossi Sharon\fnref{label1}}
 \author{Bagrat Khachatryan\fnref{label2}}
 \author{Dima Cheskis\fnref{label1}}

\fntext[label1]{the Physics Department, Ariel University, Ariel, 407000, Israel}
\fntext[label2]{the Physics Department, Technion, Haifa, 3200003, Israel}
\ead{dimach@ariel.ac.il}
% \address{Address\fnref{label3}}
% \fntext[label3]{C}


% \author{Dima Cheskis}
% 
% \address{Ariel University, Israel}

\begin{abstract}
Many modern electronic devices utilize linear Hall sensors\cite{Ramsden2006} to measure current and the magnetic field,
as well as to perform switching and latching operations\cite{Honeywell2011}.
Smartphones, laptops, and e-readers all work with very low (sub-mA) currents.
To perform a switching function in such low-power devices, however,
a Hall sensor must be able to work in the $\mu$A regime. 
This paper demonstrates, for the first time, the ability of a standard Hall detector to work in the μA regime between 0 and 0.7 Tesla. 
A second important application of this technology is the measurement of electron transport parameters in thin films,
which is essential to elucidating their electronic behavior. 
The development of new devices using thin films demands very precise measurements of tiny electrical currents, 
low-intensity magnetic fields, and other small signals. The proposed system delivers a very small but stable electric current without external noise, 
and can be used to measure small transport parameters with very high precision.
We demonstrate the capabilities of this system by measuring the slope
of the Hall effect\cite{Hurd2012b} with a four-point probe at current intensities of 100, 10, and 1 $\mu$A. 
\end{abstract}


\end{frontmatter}

% \linenumbers


%----------------------------------------------------------------------
% SECTION I: Introduction
%----------------------------------------------------------------------

\section{Introduction}

Many modern devices, such as smartphones, tablets, e-readers, GPS units,
and heart rate monitors, are controlled remotely and operated continuously for long stretches of time.
These features come with a practical design constraint: the devices spend most of their time in a low-power “sleep” mode,
using a battery or a DC low-power bus to deliver current in the sub-milliampere (mA) regime. When transmitting information,
the devices switch to high-power radio frequency (RF) activity, and the current increases to the mA or even Ampere scale.
To control these switching behaviors, precise sensors must be used that are capable of measuring both low and high currents. 
Usually, noninvasive current control is done using Hall sensors.
The current flowing through the Hall sensor creates a perpendicular Hall voltage,
which is proportional to the measured current and detectable through the material of the sensor.
Weak magnetic fields can be measured by various techniques,
such as superconducting quantum interference (SQUID) devices\cite{Beyer2008} or magnetoresistance sensors (``giant'', ``anomalous'',
or ``tunneling'', respectively denoted GMR, AMR, or TMR)\cite{Djamal2012,Zhao2013,Chen2012a}.
Hall sensors are also cheap, so are used in many devices. However, to conserve battery life,
the energy used by different types of sensors needs to be minimized. Usually, the output current of the Hall sensor is in the mA regime.
This current level gives a good signal-to-noise ratio (SNR), and its Hall voltage can be easily detected. 
In this paper, we develop a Hall sensor that can work in the microampere regime and still react while a device is in the sleep mode.
In order to achieve this goal, it is necessary to first greatly reduce SNR,
and then find an appropriate voltage sensor to detect the signal.
The present work focuses on the first factor.
We demonstrate for the first time a linear DC Hall sensor working in the microampere regime,
with a SNR similar to the mA regime used by current technologies. We measure magnetic fields in the range 0-0.7 Tesla,
for three very low DC current sources. 

\newpage
%----------------------------------------------------------------------
% SECTION II: Experimental Setup
%----------------------------------------------------------------------
\section{Noise treatment}

In electronic circuits, the main type of the noise at room temperature in DC circuits is the thermal or Johnson-Nyquist noise.
The noise level of the sensor voltage equals\cite{Jung2001} 

\begin{equation}
 V=\sqrt{4k_{b}TR\Delta f}
\end{equation}

where $k_b$ is the Boltzmann constant, T is the temperature, R is the resistance, and $\Delta f$ is the frequency bandwidth.
The Hall voltage signal is equal to
\begin{equation}
%  V_{Hall}=\dfrac{BI}{nqt}
 V_{Hall}=BI/nqt
\end{equation}

Hence the SNR is equal to

\begin{equation}
%  \frac{S}{N}=\dfrac{BI}{nqt\sqrt{4k_{b}TR\Delta f}}
  S/N=BI/(nqt\sqrt{4k_{b}TR\Delta f})
\end{equation}
                                                                            
The smallest magnetic field B that can be discerned corresponds to $\frac{S}{N}\approx 1$ .
This constraint leads to the relation\cite{Boero2003} 
\begin{equation}
%  B_{min}=\dfrac{V_{noise} nqt}{I}
 B_{min}=V_{noise}nqt/I
\end{equation}

Hence, if we want the same discriminating power in the $\mu$A regime that we now have in the mA regime,
it is necessary to reduce the noise in the voltage by three orders of magnitude.
The noise of a typical current source connected to the electrical grid is $\propto$ $\frac{\mu A}{\sqrt{Hz}}$.
Laboratory  power supplies are designed to provide currents and waveforms over a wide range of intensities and frequencies. They are therefore built with many electronic components, which increase the noise in the system. Moreover,  the alternating voltage of the power supply grid introduces additional noise. The standard way to reduce noise is to decrease the bandwidth using a modulated signal and a lock-in amplifier. In our case, we instead use a DC battery to supply a small amount of current without introducing any noise from extraneous electronic components or the external grid.
In this way we reduce the noise to $\propto$ $\frac{nA}{\sqrt{Hz}}$.
In the next section, we explain our current source in detail.

\newpage

%----------------------------------------------------------------------
% SECTION III: Current Source
%----------------------------------------------------------------------
\section{Current Source}

\begin{figure*}[h]
 \begin{center}
  \includegraphics[scale=0.1]{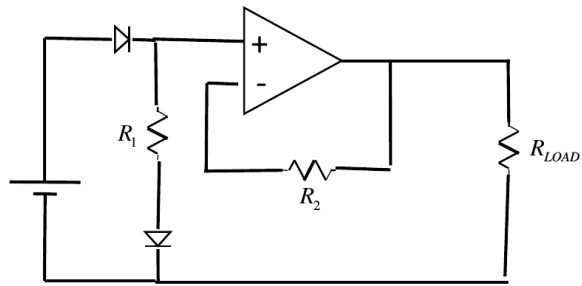}
 \end{center}
    \caption{Current source}
    \label{fig:Current source}
\end{figure*}

As was previously described, our main innovation is to use a power source with very low noise which shown in Figure \ref{fig:Current source}.
We use a 9V battery as the source of voltage, and disconnect the measurement circuit completely from the external network.
The load on the source should be no more than 9V, and ideally much less to reduce noise. 
To create a constant current through the sample, we use an operational amplifier.
If the voltage load is not large to begin with, and fluctuations in the load are small,
then the noise in the current is very small.

In order to test our sensor, we build an experiments which performs
Hall Effect measurements on an industrial GaAs sensor.  

%----------------------------------------------------------------------
% SECTION IV: Experimental Setup
%----------------------------------------------------------------------
\section{Experimental Setup}

\begin{figure}[ht]
 \begin{center}
  \includegraphics[scale=0.6]{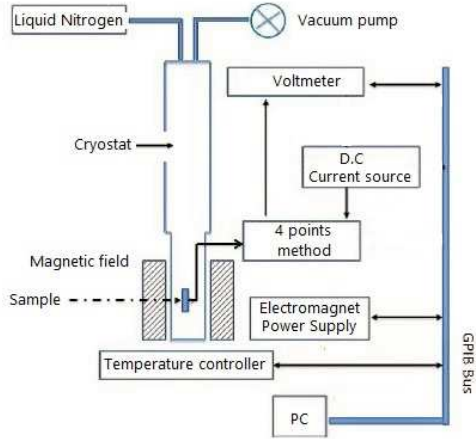}
 \end{center}
    \caption{Experimental Setup}
    \label{fig:Experimental Setup}
\end{figure}

Our setup for making Hall Effect measurements is shown in Figure \ref{fig:Experimental Setup}. This system consists of a DC current source,
a very accurate nanovoltmeter, Helmholtz magnetic coils, and a liquid nitrogen cryostat.

\begin{figure}[ht]
\begin{center}
\includegraphics[scale=0.35]{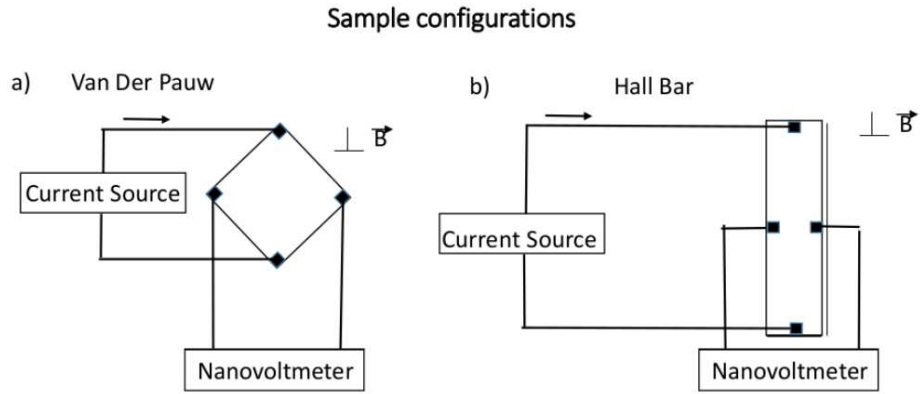}
\end{center}
    \caption{Hall measurement configurations}
    \label{fig:Hall config}
\end{figure}

With this system, we can measure both the resistance and the Hall Effect. 
This can be done using two different configurations: a Hall bar\cite{Yang2008,Yang2016} 
and the Van der Pauw sample arrangement\cite{Blanchard2000,Borup2015,Schumacher2017}, as shown in Figure \ref{fig:Hall config}. 
In the Van der Pauw arrangement, the electrical contacts are connected to the boundaries of a square sample and the Hall 
voltage is measured diagonally. 
In a Hall bar, the current flows through an elongated plate and the Hall voltage is measured at the cross section. 
In both cases, the voltage is measured perpendicular to the current, and the voltage and current contacts are separated. 
Such configurations are called “four-point probes”. 
Measurements of resistance effects in these configurations are only related to the properties of sample, 
not to the measurement circuit.

The sample is attached to a special holder with connected wires. 
The plane of the holder is perpendicular to the axis of the magnetic coils and can rotate 180 degrees. 
The holder is inserted into the evacuation area of the cryostat. 
This allows us to hold the sample either in vacuum or at a low constant gas pressure. The gas delivered to this region is pure helium. The outer walls of the cryostat are cooled by liquid nitrogen. The inside cools to 77 Kelvin, so the helium remains gaseous at a pressure of a few millibars. The temperature of the area containing the sample can be adjusted from 77K to 300K. 
This control is achieved by cooling the gas or adjusting the heater with the help of a Cernox temperature sensor 
and a controller from Lake Shore Cryotronics, Inc.\cite{lakeshore}.
This control system stabilizes the sample area to a temperature constant within 0.1 K. 
Measurements of the Hall voltage are performed under a constant magnetic field. The field can be adjusted from 0 to 0.8 Tesla, in both directions.
The sample that we use is the HSP-T Hall Sensor of the Cryomagnetic, Inc.\cite{cryomagnetics}.  
The sample type is a Hall bar with one voltage output. 
The sample is completely isolated from the external environment in the evacuation chamber of the cryostat, which makes it possible to avoid oxidation.
Although we do not know the material of the sample or its exact thickness, the manufacturer provides the dependence of the Hall voltage on the magnetic field. 
Hence, we can compare our measurements with those of the manufacturer. 
As it was shown previously, the Hall voltage depends on the magnetic field, 
the current, the charge of the carriers, the population density of the charge carriers, and the thickness of the sample.
Changes to the measured Hall voltage can only occur because the magnetic field B or electric current I change.
All other factors are constant for the sample.

\newpage

%----------------------------------------------------------------------
% SECTION V: Experimental Results
%----------------------------------------------------------------------
\section{Experimental Results}

\begin{figure}[h]
 \begin{center}
  \includegraphics[scale=0.3]{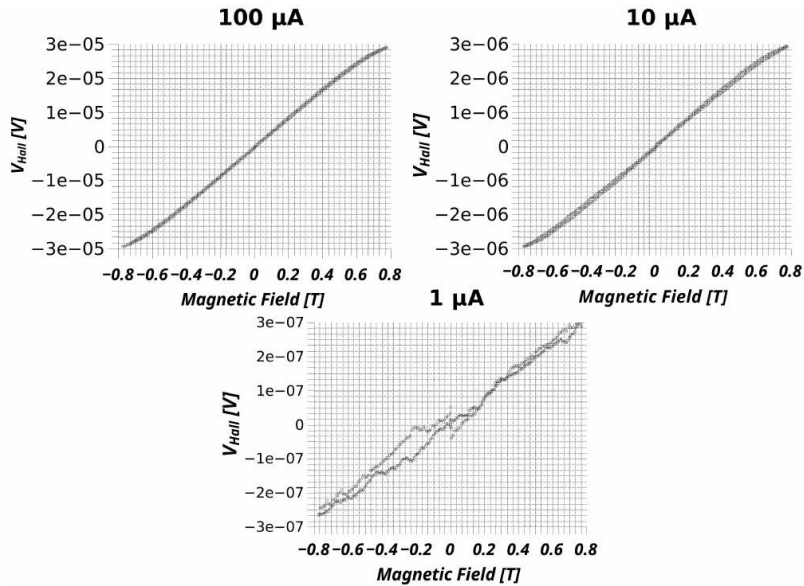}
 \end{center}
    \caption{Measured Hall voltage, $V_{Hall}$, as a function of the magnetic field for three current intensities}
    \label{fig:Hall Voltage}
\end{figure}

The Hall bar calibration sample has a linear Hall effect when the magnetic field is perpendicular
to the current passing through the sample. 
The manufacturer provides the ratio of the Hall voltage to the magnitude of 
the magnetic field for a current of 100 mA. In order to test our system and
learn its sensitivity, we measured the Hall voltage as a function of the external
field using our four-point probe. Due to our assumption that the density of carriers does 
not depend on the current, we can measure this relation for different currents and check whether
it is proportional to the manufacturer’s values. 
We measured the Hall effect under three  currents: 100 $\mu$A and 10 $\mu$A and 1 $\mu$A. 
The dependences of Hall voltage as function of magnetic field  are shown in Figure \ref{fig:Hall Voltage}.

This figure shows an approximately linear dependence of the Hall voltage on magnetic field for all current values.
In Table 1, we compare the slopes of the Hall effects measured at all currents to the slope provided by the manufacturer at 100 mA.
% The manufacturer-provided \say{sensitivity} (tolerance) of the slope was used to make comparisons
with statistical errors on the slopes derived from the least-squares linear fits.

\begin{table}[ht]
% \centering
\resizebox{\textwidth}{!}{
\begin{tabular}{|l| l| l| l| l|}
\hline
\textbf{Current} & \textbf{1$\mu$A} & \textbf{1$\mu$A}& \textbf{1$\mu$A} & \textbf{100mA (manufacturer)}\\
\hline
\textbf{Hall voltage/magnetic field}
% (proportional to current) & 0.371\tfrac{$\mu$V}{T} & 3.955\tfrac{$\mu$V}{T} & 39.85\tfrac{$\mu$V}{T} & 39.93\tfrac{$\mu$V}{T}\\
 & 0.371 $\mu$V/T & 3.955 $\mu$V/T & 39.85 $\mu$V/T & 39.93 mV/T\\
\hline
\textbf{Uncertainty on best-fit slope}
 & 1.5 nV/T & 6.5 nV/T & 81 nV/T & 80 $\mu$V/T\\
\hline
\textbf{Relative uncertainty on slope} & 419 m$\%$ & 164 m$\%$ & 205 m$\%$ & 200 m$\%$ \\
\hline
\end{tabular}
}
\caption{Hall coefficients for 1$\mu$A, 10$\mu$A, 100$\mu$A  current sources}
\end{table}

The slopes at 1 $\mu$A and 10 $\mu$A differed from the slope at 100 $\mu$A by factors of 107 and 10.07, respectively.
These deviations were very close to the expected factors of 100 and 10.
The most precisely determined slope in this series of measurements was for the 10 $\mu$A current,
which had a relative error of only 165 m$\%$. 
The uncertainties in the slopes (reported in Table 2) are directly proportional
to the root-mean-squared error (RMSE) of the Hall voltage measurements.
RMSE, measured in nanovolts, defines the real limit of our ability to resolve nonlinear phenomena in the sample.
For example, the Hall voltage curve displays hysteresis, being differently shaped for increasing and decreasing magnetic fields.
One possible source of hysteresis is a small amount of spontaneous magnetization in the sample.
The manufacturer does not emphasize the hysteresis deviation, but this effect is well known in the literature.
Figure 5 shows zoomed-in plots of the hysteresis deviation for all three currents.
The hysteresis deviations were a major source of uncertainty in the slopes.

\begin{figure}[ht]
\begin{center}
 \includegraphics[scale=0.25]{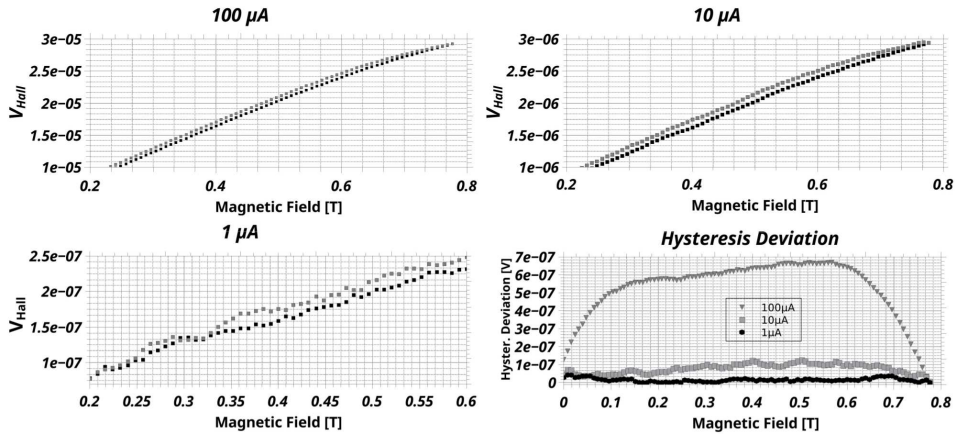}
\end{center}
    \caption{Hysteresis deviation in $V_{Hall}$ [V] for the 1,10 and 100 $\mu$A signals, as a function of the magnetic field}
    \label{fig:Average}
\end{figure}

\newpage

Table 2 reports the maximal separations in Hall voltage due to hysteresis.
The smallest hysteresis separation in absolute terms occurred at 1 $\mu$A. For all three cases,
the hysteresis deviation was larger than the precision of the voltage measurements, characterized by the RMSE.
As the current increased from 1 to 100 $\mu$A, the hysteresis deviation approached the precision limit,
but the statistical error in the slope measurement decreased.
In our experiment, the 10 $\mu$A current source offers a good compromise of these opposing effects.
% 
% \vspace{2mm}
% \begin{table}[h] 
% \begin{center} 
% \begin{tabu} to 0.8\textwidth{|c|c|c|c|}
% \hline
% Current & 1$\mu$A & 10$\mu$A & 100$\mu$A \\
% \hline
% RMSE (accuracy limit) & 10.4 nV& 48.7 nV& 530 nV \\
% \hline
% Hysteresis Deviation & 0.09 $\mu$V& 0.1 $\mu$V& 0.6 $\mu$V \\
% \hline
% \end{tabu}
% \caption{ RMSE(accuracy limit) and hysteresis deviation for 1$\mu$A, 10$\mu$A, 100$\mu$A  current sources}
% \end{center}
% \end{table}
% \vspace{2mm}

\begin{table}[h]
\centering
\resizebox{0.7\textwidth}{!}{
\begin{tabular}{|l| l| l| l|}
\hline
\textbf{Current} & \textbf{1$\mu$A} & \textbf{10$\mu$A}& \textbf{100$\mu$A} \\
\hline
\textbf{RMSE (accuracy limit)}
% (proportional to current) & 0.371\tfrac{$\mu$V}{T} & 3.955\tfrac{$\mu$V}{T} & 39.85\tfrac{$\mu$V}{T} & 39.93\tfrac{$\mu$V}{T}\\
 & 10.4 nV& 48.7 nV& 530 nV \\
\hline
\textbf{Hysteresis Deviation}
 & 0.09 $\mu$V& 0.1 $\mu$V& 0.6 $\mu$V \\
\hline
\end{tabular}
}
\caption{RMSE(accuracy limit) and hysteresis deviation for 1$\mu$A, 10$\mu$A, 100$\mu$A  current sources}
\end{table}
%----------------------------------------------------------------------
% SECTION VI: Conclusions
%----------------------------------------------------------------------
\newpage
\section{Conclusion}
We have shown that a standard industrial Hall sensor exhibits a linear response for currents in the microampere range.
We were able to deliver stable currents in the μA regume by using a modified electric circuit free
of extraneous electrical components and the external power grid. We characterized the precision of
the Hall effect slope measurement for three different current intensities, and measured hysteresis in the Hall
voltage at these currents. We find that the hysteresis significantly affects the precision of the Hall effect slope measurement.
There is a trade-off between two sources of uncertainty: for smaller currents,
the hysteresis deviation in the measured voltage is smaller, so hysteresis has less impact on the Hall effect.
On the other hand, the RMSE of individual voltage measurements increases and becomes comparable to the hysteresis.
In our setup, a current intensity of 10 $\mu$A offers the best compromise between these effects, yielding
the most precise measurement of the slope. The hysteresis deviation is a nonlinear function of current,
and decreases more slowly than the RMSE increases. This result confirms that the hysteresis deviation originates
from internal magnetic effects and not from statistical noise. 
These interesting results are leading us to design additional low-noise current sources that can further decrease
the RMSE of individual voltage measurements while preserving Hall linearity. The long-term goal of
this research is to develop low-cost, accessible methods to accurately characterize the electronic
behavior of thin films. Modern technological devices offer many advantages, but encounter new power
management problems that can only be solved by measuring low-intensity magnetic fields with stable and simple systems.
Our experiment shows a new way to understand and measure such small effects, and may eventually lead
to the development of new equipment that can take full advantage of currents in the microampere regime.

%----------------------------------------------------------------------
% SECTION VII: Aknowledgment
%----------------------------------------------------------------------

\section*{Acknowledgment}

We would like to thank to Prof. Boris Ashkinadze from Technion, Haifa, Israel, for his advise and expertise in the area of transport measurements. 
This study was partialy founded by Kamin program of Israel Innnovation Authority.

\bibliographystyle{elsarticle-num}
% \bibliography{Instrumentation_paper,Cryomagnetics,Lakeshore}
\bibliography{Instrumentation_paper}

\end{document}